\documentclass[pra,12pt,english]{article}
\input epsf

\begin{document}
\title{Protective Measurements}

\maketitle

\vspace {.1cm}

Protective measurement \cite{AVPM} is a method for measuring an
expectation value of an observable on a single quantum system. The
quantum state of the system can be protected by a potential, when
the state is a nondegenerate energy eigenstate with a known gap to
neighboring states, or via Zeno effect by frequent projection
measurements.

Apart from protection, the  procedure consists of  a standard von
Neumann measurement  with weak coupling  which is switched on and,
after a long time, switched off, adiabatically. The interaction
Hamiltonian for protective measurement of $O$ is:
\begin{equation}\label{PM}
  H_{int} = g(t) P O,
\end{equation}
where $P$ is a momentum conjugate to $Q$, the pointer variable of
the measuring device. The interaction Hamiltonian is small as  in
{\it weak measurements} \cite{WV}.  In both cases the initial state
of the pointer is such that $\langle Q \rangle_{in} =0,~~\langle P
\rangle_{in} =0$. In weak measurement, the weakness is due to small
uncertainty in $P$ which requires a large uncertainty of the pointer
variable $Q$. Thus,  although for the final wave function of the
pointer, $\langle Q \rangle_{fin}=\langle \Psi | O |\Psi \rangle$, a
single measurement does not allow obtaining significant information
about $\langle \Psi | O |\Psi \rangle$. In protective measurement,
the pointer is well localized at zero, which requires large
uncertainty in $P$ and the weakness is due to a small value of the
coupling $g(t)$. The coupling to the measurement device is weak, yet
long enough  so that we still have $\int g(t) dt =1$. The result is
again $\langle Q \rangle_{fin}=\langle \Psi | O |\Psi \rangle$, but
this time, the pointer is well localized, so we can learn the value
of the expectation value from a single experiment. This is so if
during the measurement, the quantum state of the system remains
close to $|\Psi \rangle$. Given the adiabatic switching of the
measurement interaction, its small value, and the protection of the
state, this is indeed the case.

One of the basic results of quantum mechanics is that when a
measurement of a variable $O$ with eigenvalues $o_i$ is performed on
a quantum system described by the state $|\Psi \rangle$, the
probabilities $p_i$ for obtaining outcome $o_i$ satisfy:
\begin{equation}\label{ExV}
\langle \Psi | O |\Psi \rangle=\sum p_i o_i .
\end{equation}
This is why the expression $\langle \Psi | O |\Psi \rangle$ is
called the expectation value of $O$. In protective measurements we
obtain this value not as a statistical average, but as a reading of
a measuring device coupled to a {\it single} system.

A sufficient number of protective measurements performed on a single
system allow measuring its quantum wave function. This provides an
argument against the claim that the quantum wave function has a
physical meaning only for an ensemble of identical systems.
Therefore, protective measurements have some merit even when the
protection is achieved via frequent projection measurements on the
state $|\Psi \rangle$ with no new information  obtained during the
whole procedure. If the protection of the state is via a known
energy gap to any orthogonal state, then the protection measurement
provides new information: we can find the whole wave function. Thus,
protective measurement of the quantum wave function of an ion in a
trap can  yield the the trap's potential.

Numerous objections to the validity and meaning of protective
measurements have been raised \cite{Ob2,Ob3,Ob5,Ob6,Ob7}. The
validity of the result was questioned due to misunderstanding of
what the protective measurement is \cite{AAVMPM,ROY,CrPM}. The issue
of meaning: ``Is the wave function of a single particle an
ontological entity?'' \cite{AAVPM} is open to various
interpretations. Some will say `yes' even before hearing about
protective measurement, others say `no' just because protective
measurements are never 100\% reliable. The protective measurement
procedure is not a proof that we should adopt one interpretation
instead of the other, but it is a good testbed which shows
advantages and disadvantages of various interpretations. For
example, the Bohmian interpretation does not provide a natural
explanation of how   a protective measurement can ``draw'' the whole
wave function of an ion in a ground state of a trap, since the
Bohmian position of the ion   hardly changes during the measurement
\cite{AVBohm,AES}.

The protective measurements method can be extended to  pre- and
post-selected systems described by a two-state vector $\langle{\Phi}
\vert ~~\vert\Psi\rangle$ \cite{TSVF}. It requires separate
different protections for the forward and backward evolving quantum
states which are achieved by pre- and post-selection of quantum
states of systems which provide the protection \cite{AV2SP}. The
outcome of such protective measurements is not the expectation
value, but the {\it weak value}, ${ \langle{\Phi} \vert O
\vert\Psi\rangle \over \langle{\Phi}\vert{\Psi}\rangle }$ \cite{WV}.
A realistic setup for such protective measurement is a weak coupling
to a variable of a decaying system which is post-selected not to
decay \cite{Decay}.

Theoretical analysis of protective measurements leads to deeper
understanding of quantum reality while its experimental realization
(which seems feasible in a near future) might be useful for more
effective gathering of information about quantum systems \cite{Nus}.

This work has been supported in part by the European Commission
under the Integrated Project Qubit Applications (QAP) funded by the
IST directorate as Contract Number 015848 and by grant 990/06 of the
Israel Science Foundation.

\vskip .5cm Lev Vaidman\hfill\break The Raymond and Beverly Sackler
School of Physics and Astronomy\hfill\break Tel-Aviv University,
Tel-Aviv 69978, Israel
\end{document}